\documentstyle[12pt,psfig]{article}
\begin{document}

\title{A Hartree-Fock {\em ab initio} band-structure calculation
employing Wannier-type orbitals}

\author{Martin Albrecht, Alok Shukla, Michael Dolg, Peter Fulde, \\
{\protect\it Max-Planck-Institut f\"{u}r Physik komplexer Systeme}, \\
{\it N\"{o}thnitzer Str. 38, D-01187 Dresden, Germany } \\[5mm]
Hermann Stoll \\
{\protect\it Universit\"{a}t Stuttgart}, \\
{\it Institut f\"{u}r Theoretische Chemie, D-70550 Stuttgart, Germany } \\[5mm]}

\maketitle

\newpage 
\begin{abstract}
\hspace*{-12.3pt} An {\em ab initio} Wannier-function-based approach
to electronic ground-state calculations for crystalline solids is outlined. 
In the framework of the linear combination of atomic orbitals method 
the infinite character of the solid is rigorously taken into
account. The Hartree-Fock ground-state energy, cohesive energy, lattice constant
and bulk modulus are calculated in a fully {\em ab initio} manner as it is
demonstrated for sodium chloride, NaCl, using basis sets close to the
Hartree-Fock limit. It is demonstrated that the Hartree-Fock band-structure
can easily be recovered with the current approach and agrees with the one
obtained from a more conventional Bloch-orbital-based calculation.
It is argued that the advantage of the present approach lies in its
capability to include electron correlation effects for crystalline 
insulators by means of well-established quantum chemical procedures.
\end{abstract}
%
\section{Introduction}
\label{intro}
Calculations of ground-state properties as well as band-structure
calculations
are usually done in the so-called Bloch-orbital approach that employs
itinerant orbitals, which are delocalized all over the infinite crystal\cite{crystalbook}.
The wavefunctions obtained, in a single-determinantal framework, naturally fall into
irreducible representations according to 
the translational symmetry of the system.
Thus this method is well adapted to the problem at hand. The state-of-the-art
of Hartree-Fock (HF) calculations following this approach is represented by the
CRYSTAL program\cite{crystalprog}. On the other hand,
to go beyond the HF approach means to allow for electron correlation which
is a more or less local phenomenon. Here, the use
of delocalized Bloch orbitals might not be the optimum choice.

Of course, the latter remark only applies to wavefunction-based methods 
aimed at explicitly including electron correlation at a microscopic level.
In contrast to that, density-functional theory (DFT) implicitly incorporates 
correlation into a local potential/energy-density, formally still at the independent-particle
level\cite{lda}. Very often DFT band-structures are, without rigorous theoretical justification,
directly identified with the eigenvalue spectrum of the Kohn-Sham (KS) equations. 
A better theoretical foundation has been achieved using 
the GW approximation which has been developed for calculating quasi-particle excitations
in semiconductors and insulators\cite{hott}, but still the approximations are not fully
controlled.

Wavefunction-based methods for solids suggested in literature so far are either
of the Quantum-Monte-Carlo\cite{QMC} or configuration-interaction (CI) type.
In the latter case, one tries to explicitly exploit the local nature of the
correlations by drawing advantage of the highly developed and accurate 
{\em ab initio} quantum-chemical methods\cite{fuldebook}. 
Various kinds of such a 'local ansatz' in the framework of nonorthogonal
orbitals have been proposed by Stollhoff and Fulde \cite{stolh} almost
two decades ago and yielded a considerable amount of useful results since then. 
More recently the scheme  of local increments\cite{stoll92} was formulated 
using the computationally more attractive orthogonal orbital sets as well as
standard quantum chemical methods like coupled-cluster (CC). The scheme has been
applied quite successfully for the determination of ground-state
properties of a large variety of systems, including elementary\cite{paulus95}
and III-V semiconductors\cite{paulus96} as well as ionic compounds\cite{doll95}.
First successful applications to correlated band-structure calculations
for valence bands have also been reported\cite{juergen1,juergen2}. However,
the main drawback of the incremental scheme, both for ground-state
properties as well as band-structures, is the derivation of the correlation
increments from finite-cluster instead of infinite-solid calculations.

By consistently and rigorously taking into account the periodicity of the 
lattice, at all stages of the calculations, we now strive to provide an
improved procedure. In two recent papers\cite{alok1,alok2}, henceforth denoted 
as I and II, we demonstrated the possibility of calculating ground-state
properties of ionic solids by a
procedure which has the novel feature that it directly yields localized HF
orbitals, as the result of a self-consistent-field (SCF) calculation
in the linear combination of atomic orbitals (LCAO) framework. In
this paper, we extend our considerations in a two-fold way. In the first place,
we focus on the Hartree-Fock {\em ab initio} band-structure.
Secondly, we enhance the numerical stability of our approach, so that
calculations of localized orbitals become also possible for extended 
state-of-the-art basis sets approaching the HF limit.
The localized orbitals at hand, we plan to include correlations in future
investigations by means of the incremental scheme.

This paper is organized as follows. In Section \ref{hft} we briefly repeat
the theory within a restricted Hartree-Fock (RHF) framework. A
more detailed description is given in paper II. Section \ref{bs}
gives  the formulae needed to obtain the band-structure by our approach.
Numerical results for the ground-state energy, the lattice constant,
the bulk modulus and the band-structure of NaCl are presented in Section \ref{results}. 
Finally, Section \ref{conclusion} contains the conclusions.

\section{Theory}
\label{theory}
\subsection{Hartree-Fock Equations}
\label{hft}
Our aim is to directly determine localized (Wannier-type) orbitals
for the solid.  Denoting by $|\alpha({\bf R}_j)>$ the Wannier orbitals of a unit cell
located at lattice vector ${\bf R}_{j}$,
the set $\{ |\alpha({\bf R}_j)>; \alpha =1,n_{c}; j=1,N \}$  
spans the occupied HF space of the solid. (Here, $n_c$ is the number of orbitals
per unit cell, and  $N (\rightarrow \infty)$ is 
the total number of unit cells in the solid.)
Translational symmetry requires
\begin{equation}
|\alpha({\bf R}_i)> = {\cal T}({\bf R}_i) 
|\alpha({\bf 0})> \mbox{,}
\label{eq-trsym}
\end{equation}
where ${\cal T}({\bf R}_i)$ is an operator which represents a translation
by the lattice vector ${\bf R}_{i}$, and $|\alpha({\bf 0})>$ denotes
the orbitals of an (arbitrarily chosen) reference cell with lattice vector ${\bf R = 0}$.
Assuming now the $|\alpha({\bf R}_j)>$ to be orthogonal:
\begin{equation}
<\alpha({\bf 0}) | {\cal T}({\bf R}_k) | \beta ({\bf 0}) > = \delta_{\alpha\beta} \delta_{0k}
\mbox{,}
\label{ortho}
\end{equation}
the closed-shell HF energy (per unit cell) for the solid can be written as
\begin{eqnarray}
E      &    =&    2 \sum_{\alpha=1}^{n_{c}} <\alpha ({\bf 0}) |T|\alpha ({\bf 0}) > +
   2 \sum_{\alpha=1}^{n_{c}} <\alpha ({\bf 0}) |U|\alpha ({\bf 0}) >   \nonumber  \\ 
        &    +  & \sum_{\alpha,\beta=1}^{n_{c}} \sum_{j=1}^{N}
 ( 2 <\alpha ({\bf 0}) \beta ({\bf R}_j)|\alpha ({\bf 0}) \beta ({\bf R}_j)>
 - <\alpha ({\bf 0}) \beta ({\bf R}_j)|\beta ({\bf R}_j)\alpha ({\bf 0}) >)  
\nonumber  \\
 &  + & E_{nuc} \mbox{.} 
\label{eq-esolidf}
\end{eqnarray}
Here, $T$ and $U$ denote the 
kinetic-energy and  electron-nucleus potential-energy operators, respectively,
$<\alpha \beta |\alpha \beta>$ and $<\alpha \beta |\beta\alpha >$ represent the usual
 two-electron Coulomb and exchange integrals, and $E_{nuc}$ is the interaction
energy (per unit cell) of the nuclei. The divergent averages of $U$, $E_{nuc}$, and
the Coulomb term are already assumed to have been eliminated from Eq.\ 
\ref{eq-esolidf}.

Minimization of the $|\alpha ({\bf R}_j)>$, under the orthogonality condition, Eq.\
\ref{ortho}, can be shown to lead to the following (modified) Hartree-Fock equations
\begin{equation}
( T + U
 + \sum_{\beta} (2  J_{\beta} -  K_{\beta})   
+\sum_{k \in{\cal N}} \sum_{\gamma} \lambda_{\gamma}^{k} 
|\gamma ({\bf R}_k)>
<\gamma ({\bf R}_k)| ) |\alpha ({\bf 0})>
 = \epsilon_{\alpha} |\alpha ({\bf 0})>
\mbox{,}
\label{eq-hff1}         
\end{equation}  
to be solved in the limit $\lambda_{\gamma}^{k}  \rightarrow \infty $. 
The results are not sensitive to the actual values of the $\lambda_{\gamma}^{k} $, as long as they are large
enough; in our calculations, we chose $\lambda_{\gamma}^{k} = 10^4 H$.
$J_{\beta} $ and $K_{\beta}$ are the conventional Coulomb and exchange operators
\begin{equation}
\left.
 \begin{array}{lll}
 J_{\beta}|\alpha> & = & \sum_{j} <\beta ({\bf R}_j)|\frac{1}{r_{12}}|
\beta ({\bf R}_j)>|\alpha> \\  
 K_{\beta}|\alpha> & = & \sum_{j} <\beta ({\bf R}_j)|\frac{1}{r_{12}}|\alpha>
| \beta ({\bf R}_j)> \\  
\end{array}
 \right\}  \mbox{;} \label{eq-jk} 
\end{equation}
as discussed in the earlier papers I,II,\, $U,J,K$  involve infinite lattice sums which are 
rigorously performed, in our treatment, apart from suitable truncation criteria.
For solving Eq.\ \ref{eq-hff1} self-consistently, we employ a finite-basis-set LCAO expansion, 
with the basis set describing the $|\alpha ({\bf 0})>$ extending over all atoms of a 
neighborhood ${\cal N}$ of the reference unit cell. ${\cal N}$ should consist of all unit cells with basis
functions having non-negligible overlap to those of the reference cell;
clearly, the choice of ${\cal N}$ 
will also be dictated by the system under consideration---the more delocalized
the  electrons of the system are, the larger ${\cal N}$ will need to be. 
In our calculations we have typically chosen ${\cal N}$ to include
up to third-nearest neighbor unit cells.

\subsection{Band Structure}
\label{bs}
After solving the HF equations of the solid for localized Wannier-type orbitals $|\alpha({\bf R}_j)>$,
we have at our disposal, among other results, the converged Fock matrix in the AO basis:
\begin{equation}
F_{p0,qj} = <p({\bf 0}) |  T + U
 + \sum_{\beta} (2  J_{\beta} -  K_{\beta}) | q({\bf R}_j) >   
\mbox{.}
\label{eq-band1}         
\end{equation}  
(The matrix elements are easily obtained from Eq.\ \ref{eq-hff1}, by subtracting off the last (one-electron) term.)

The band-structure involves matrix elements of the Fock operator in $\bf k$-space.
The $\bf k$-space transform of the AOs $|p({\bf R}_j)>$ is
\begin{equation}
  |p({\bf k})>=\frac{1}{\sqrt{N}} \sum_{{\bf R}_{\rm j}}{\rm
  e}^{{\rm -i}{\bf k}{\bf R}_{\rm j}}
  |p({\bf R}_j)>.
\label{eq:kversusr}
\end{equation}
This immediately renders the transformation rule for operators:
\begin{equation}
Q_{pq}({\bf k})=\sum_{{\bf R}_{\rm j}}{\rm e}^{-{\rm i}{\bf k}{\bf R}_{\rm
j}}
\left<p({\bf 0})| Q | q({\bf R}_j) \right>.
\label{eq:otraf}
\end{equation}
Inserting $Q=F$ (from Eq.\ \ref{eq-band1}) and $Q=1$, leads to the ${\bf k}$-space Fock and overlap
matrices, $F_{pq}({\bf k})$ and $S_{pq}({\bf k})$, respectively. A final diagonalization in ${\bf k}$-space,
i.e.\ solving the generalized eigenvalue equation
\begin{equation}
\sum_{q} F_{pq}({\bf k}) C_{q\alpha}({\bf k})  =  \epsilon_{\alpha}({\bf k})  \sum_{q} S_{pq}({\bf k})  C_{q\alpha}({\bf k})
\end{equation}
yields the desired quasiparticle energies for a given $\bf k$ point.


%
%
%
%
%
\section{Calculations and results}
\label{results}
We have performed an all-electron calculation for the NaCl crystal adopting
the experimental face-centered cubic (fcc) structure. The unit cell was
defined with sodium at the $(0,0,0)$ position and chlorine at 
$(0,0,a/2)$, where $a$ is the lattice constant.
In the following, we will repeatedly compare our
results to those obtained using the Bloch-orbital-oriented approach of the Torino group reported in~\cite{torino1} for
the ground-state properties. For the band-structure, we performed
calculations using the most recent version CRYSTAL95 of the Torino
group~\cite{crystalprog}. In order to establish a strict comparison, 
we used their most recent basis sets\cite{torino1}, i.e.\ (15s7p)/[4s3p] for Na and (19s11p)/[5s4p] for Cl.
The $p$ functions were constructed from Gaussian lobes\cite{whitten}
which yield an approximate form of Cartesian-type basis functions. The
number of basis functions per unit cell was 30. To satisfy the
orthogonality requirement between wavefunctions located in
neighboring cells, a neighborhood of 42 unit cells was taken into
account according to the previous section, which corresponds to up to
third-nearest neighbors in the fcc crystal. Thus the number of basis
functions associated with the Wannier orbitals is as large as 1290.
Note, however, that exploiting equivalence of integrals connected
by lattice translations effectively reduces the computational effort for setting up the
Fock matrix so that it is {\em not} larger than in a corresponding Bloch-orbital-based approach.

In order to establish a firm basis for a comparison of total energies
obtained within the Bloch-orbital-based approach used in CRYSTAL95
and our present method using Wannier functions we had to guarantee
that the Coulomb and exchange series in both programs are evaluated
with the same accuracy. The program CRYSTAL95 offers several input
parameters for this purpose. In the current work we chose for
ITOL1, ITOL3, ITOL4 and ITOL5 the values of 7, 7, 7 and 15, respectively,
and corresponding thresholds in our program.
This combination of parameters is believed to give well-converged results
and ensures an accuracy of $\approx 1.0\times10^{-7}$ atomic units (a.u.)
in the Fock matrix elements. The resulting accuracy of the total energy
is $\approx$ 1 milliHartree per atom~\cite{crystalprog}. In fact, at
the HF equilibrium lattice constant the total energies obtained with 
CRYSTAL95 and our approach agree within one mH.
A similar good agreement was also observed earlier for other systems
(cf. I and II).

For the derivation of the ground-state properties of NaCl,
we calculated the energy per unit cell for various lattice parameters
near the equilibrium value (table~\ref{tab-gkvar}).
We fitted these data points to a cubic polynomial to obtain
the bulk modulus, the equilibrium lattice constant and the equilibrium
energy. Our results are listed in 
table~\ref{tab-groundstate} together with the corresponding numbers obtained 
by the Torino group with their CRYSTAL program package and the
experimental values~\cite{torino1}. The lattice energy was calculated with
respect to the separated Na$^+$ and Cl$^-$ ions. To make the comparison with
the results of the Torino group meaningful, we used the same HF
energies of -161.67001 a.u. (Na$^+$) and -459.54320 a.u. (Cl$^-$), which
can be easily reproduced using the basis set reported by them~\cite{torino1}. 
As can be seen in table~\ref{tab-groundstate}, the agreement of
our results for the lattice constant and the cohesive energy with the
values obtained by the Torino group is excellent. For the bulk modulus
the disagreement is less than 1~GPa and might at least partially be attributed 
to the use of lobe vs.\ cartesian basis functions in the two approaches,
as will be discussed below.
 
\begin{figure}
\centerline{\psfig{figure=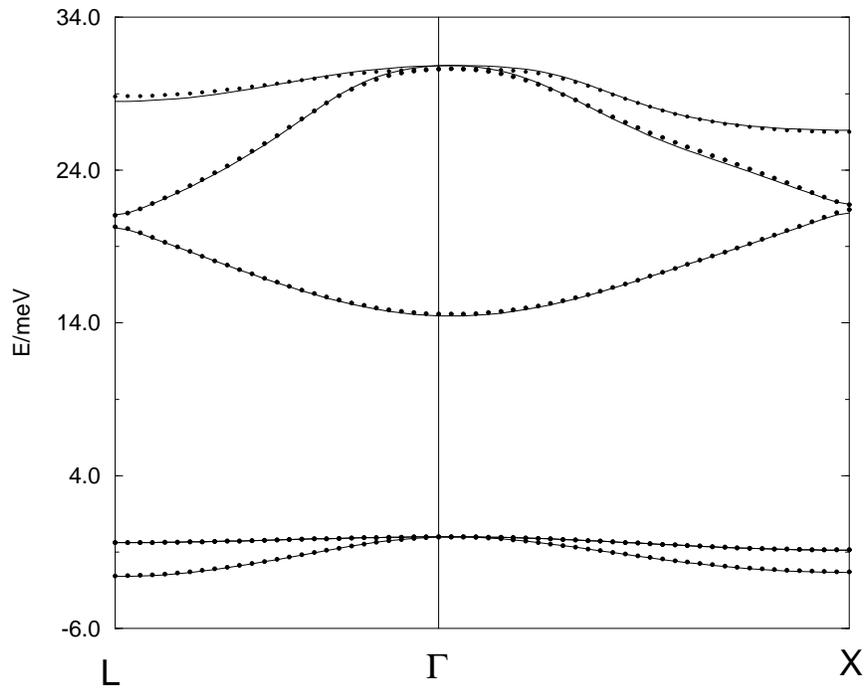,width=13cm,angle=-90}}
\caption{{\it Hartree-Fock band structure for NaCl. The highest three valence bands
(the upper one being degenerate) as well as the lowest three conduction bands are displayed. The solid line
shows results of the present work. For comparison, the results obtained by CRYSTAL95 are shown
as dotted lines. }}
\label{fig-bd}
\end{figure}

The band-structure, evaluated at the experimental lattice constant of
$ a = 10.526 au$,  is shown in Fig.~\ref{fig-bd} for two chosen high symmetry
directions in the first Brillouin zone. Displayed are the highest
three valence bands (the upper one being degenerate) as well as the lowest
three conduction bands. The zero of energy was chosen to coincide with the
highest valence point. For the sake of comparison, results obtained by
using the program package CRYSTAL95 of the Torino group with the same
basis set are plotted as dotted curve together with the results of the
present work (solid line). We observe an excellent overall agreement when 
comparing the curves obtained by the two approaches. For the sake of 
transparency we also give numerical values of the quasiparticle energies at the specified symmetry 
points in table~\ref{tab-points}.

As for the deviations between our results and the ones obtained with CRYSTAL95, 
a remark concerning a technical detail is in order here. 
As can be seen from table~\ref{tab-points}, deviations between
the two approaches are $\leq$2 mH for the valence bands,
whereas for the conduction bands differences of up to some ten mH arise
for the highest band at the L point, which is well outside the
overall agreement in the total energy per unit cell given above. This can be
traced back to the usage of Gaussian lobe basis functions in our calculations. 
To confirm this point we made test calculations on the NaCl molecule comparing 
results with cartesian Gaussian and Gaussian lobe basis functions.
The comparison of the single-particle energies reveals the
same features as in the case of the solid. A deviation of some tenth
of a mH for the highest occupied states is contrasted by a deviation of up to
13 mH in the excited states. This helps to clarify the slight deviations in
table~\ref{tab-points}. We want to point out, however, that the use of 
Gaussian lobe functions does not imply any approximations in our HF scheme, 
but merely corresponds to the use of alternative basis functions.

\section{Conclusions}
\label{conclusion}
An {\em ab initio} Hartree-Fock approach for insulating  crystals yielding 
directly orbitals in a chemically intuitive localized representation 
has been applied to band-structure calculations as well as the determination of
the lattice constant, cohesive energy and bulk modulus of sodium chloride NaCl. 
The close agreement between the results obtained using the present approach, 
and corresponding ones obtained using the conventional Bloch-orbital-based HF 
approach, demonstrates that the two schemes are entirely equivalent. 
The advantage of our method is that by taking into account local excitations 
from the Hartree-Fock reference state
using conventional quantum-chemical methods, one can go beyond the mean-field
level and study the influence of electron correlations on, e.g., on the binding 
energy and the band-structure of a solid, in an entirely {\em ab initio} manner.
Work along these lines is currently underway in our laboratory.

\newpage

\begin{table}  
 \protect\caption{Total energies (per unit cell) obtained using
our approach for NaCl,    for different
values of lattice constants.
Lattice constants are in
units of \AA, and energies are in atomic units.}
\vspace{5mm}
\protect\begin{center}   
\begin{tabular}{|cc|} \hline 
              &           \\   
 Lattice Constant & Total Energy  \\  
\hline 
   5.65       &-621.4930916\\
   5.693      &-621.4934776  \\
   5.736      &-621.4936939  \\  
   5.779      &-621.4937934  \\
   5.822      &-621.4935761           \\
   5.865      &-621.4933206       \\   
   5.95       &-621.4929614          \\
\hline  
\end{tabular}                      
\end{center}  
\label{tab-gkvar} 
\end{table}

\begin{table}  
 \protect\caption{ Equilibrium lattice constants (in \AA), bulk moduli 
(in GPa), equilibrium energies (in atomic units), and lattice energies 
(in kcal/mol) for NaCl, obtained using our approach and the one  
reported by the Torino group\protect\cite{torino1}. Relevant experimental data
are also given for comparison.}
\vspace{5mm}
\protect\begin{center}  
\begin{tabular}{|l|ll|} \hline 
           &           &         \\
 Quantity  & Method    & Results \\  
\hline
           & This Work & 5.785               \\
Lattice Constant
           & Torino   &  5.80           \\
           & Exp       & 5.57$^a$         \\ 
\hline
           & This Work & 23.15                     \\
Bulk Modulus
           & Torino   &  22.3                 \\
           & Exp       & 28.6$^b$                  \\ \hline
           & This work & -621.494    \\   
Equilibrium Energy
           & Torino   &  -621.495           \\ \hline
           & This Work &  176.1                  \\
Lattice Energy
           & Torino   &  176.1                 \\
           & Exp       & 189.2$^c$      \\ 
\hline 
\end{tabular}                      
\end{center}  
\label{tab-eqprop}    
$^a$ Ref.~\cite{gkexp}. \\
$^b$ Ref.~\cite{bulkexp}. \\
$^c$ Ref.~\cite{enexp}. \\
\label{tab-groundstate} 
\end{table}  
\begin{table}
\protect\caption{{\it Comparison of the quasiparticle energies at high symmetry
points for the first three occupied and unoccupied bands. Row ``a'' contains
the values given by CRYSTAL95, row ``b'' the corresponding values of this work.
 The energies at the gamma-point are chosen to be zero. All results
 are in atomic units}}
\vspace{5mm}
\begin{center}
\begin{tabular}{|llcccccc|}
\hline 
 & &     &     &     &     &     &     \\
 & & $1$ & $2$ & $3$ & $4$ & $5$ & $6$ \\ 
\hline
L  & a &
    -0.094  &-0.014   &-0.014    &0.746   &
     0.773 & 1.096  \\
  & b &
     -0.095&-0.014    &-0.014     &0.743   &
      0.774&1.047  \\ \hline
$\Gamma$  & a &
      0.000& 0.000   & 0.000   & 0.536  &
      1.126& 1.126   \\
  & b &
    0.000  & 0.000 & 0.000  & 0.531  &
      1.133&1.133    \\ \hline
X    & a &
     -0.084&-0.031   &-0.031    & 0.787  &
     0.799&  0.974 \\ 
    & b &
     -0.086& -0.032   &-0.032     &0.779     &
     0.800  &  0.978    \\  
\hline
\end{tabular}
\end{center}
\label{tab-points}
\end{table}


\end{document}